# Automatic Analysis System of Calcaneus Radiograph: Rotation-Invariant Landmark Detection for Calcaneal Angle Measurement, Fracture Identification and Fracture Region Segmentation

Jia Guo[1], Yuxuan Mu[1], Dong Xue[2], Huiqi Li[1]*, Junxian Chen[1], Huanxin Yan[3], Hailin Xu[4], Wei Wang[2]*

[1]Beijing Institute of Technology, Beijing 100081, China
[2]The First Affiliated Hospital of Jinzhou Medical University, Jinzhou 121001, China
[3]Zhejiang University of Science & Technology, Zhejiang 310032, China
[4]Peking University People's Hospital, Beijing 100044, China

## Abstract

*Background and Objective*: Calcaneus is the largest tarsal bone to withstand the daily stresses of weight-bearing. The calcaneal fracture is the most common type in the tarsal bone fractures. After a fracture is suspected, plain radiographs should be taken first. Bohler's Angle (BA) and Critical Angle of Gissane (CAG), measured by four anatomic landmarks in lateral foot radiograph, can guide fracture diagnosis and facilitate operative recovery of the fractured calcaneus. This study aims to develop an analysis system that can automatically locate four anatomic landmarks, measure BA and CAG for fracture assessment, identify fractured calcaneus, and segment fractured regions.

*Methods*: For landmark detection, we proposed a coarse-to-fine Rotation-Invariant Regression-Voting (RIRV) landmark detection method based on regressive Multi-Layer Perceptron (MLP) and Scale Invariant Feature Transform (SIFT) patch descriptor, which solves the problem of fickle rotation of calcaneus. By implementing

*Huiqi Li (E-mail: huiqili@bit.edu.cn) and Wei Wang (E-mail: weiwang_ly@126.com) are the corresponding authors

a novel normalization approach, the RIRV method is explicitly rotation-invariance comparing with traditional regressive methods. For fracture identification and segmentation, a convolution neural network (CNN) based on U-Net with auxiliary classification head (U-Net-CH) is designed. The input ROIs of the CNN are normalized by detected landmarks to uniform view, orientation, and scale. The advantage of this approach is the multi-task learning that combines classification and segmentation.

*Results*: Our system can accurately measure BA and CAG with a mean angle error of 3.8° and 6.2° respectively. For fracture identification and fracture region segmentation, our system presents good performance with an F1-score of 96.55%, recall of 94.99%, and segmentation IoU-score of 0.586.

*Conclusion*: A powerful calcaneal radiograph analysis system including anatomical angles measurement, fracture identification, and fracture segmentation can be built. The proposed analysis system can aid orthopedists to improve the efficiency and accuracy of calcaneus fracture diagnosis.

# 1. Introduction

Calcaneus, also known as heel bone, is the largest tarsal bone to withstand the daily stresses of weight-bearing. The calcaneal fracture is the most common type in the tarsal bone fractures, accounting for 2% of all fractures and 60% of tarsal bone fracture [1]. Imaging of a suspected fractured calcaneus usually begins with plain X-ray radiographs. If plain radiographs are not conclusive, further investigations with magnetic resonance imaging (MRI), computed tomography (CT), or nuclear medicine bone scan are required for diagnosis. Though CT represents a promising tool for surgery decisions, plain X-ray radiograph is a better screening method due to its low-cost, convenience, and less exposure to beams.

Calcaneal fractures can be divided broadly into two types: intra-articular and extra-articular, depending on whether there is articular involvement of the subtalar joint [1]. Studies have suggested that in intra-articular calcaneus fracture treatment, anatomical structures of the calcaneus can be restored validly and get to better functional recovery if the surgery is performed correctly [2][3]. Bohler's Angle (BA) [4] of the calcaneus has been used since 1931 to aid operative restoration of the fractured calcaneus and fracture diagnosis. Another angle that has been used to diagnose calcaneus fractures is the Critical Angle of Gissane (CAG) [5]. There is overwhelming evidence that restoring Bohler's angle to near normal after fracture indicates better recovery [6][7][8]. The normal range for BA and CAG in the atraumatic adult calcaneus has been quoted between $20 - 45°$ and $90 - 150°$ respectively [4]. The measurement of BA and CAG can be calculated by the location of four anatomical landmarks $L_1$, $L_2$, $L_3$ and $L_4$ as shown in Figure 1. Because BA and CAG as assessment indicators are not conclusive enough for fracture diagnosis [9], visual inspection of the fracture region in plain radiographs is needed for calcaneus fracture identification. Manual assessment of a calcaneus radiograph usually includes landmark annotation for calcaneal angle measurement, fracture judgment, and fracture region annotation (if identified as fractured), which is exhaustive and time-consuming. Therefore, an automatic analysis system of calcaneus radiograph can assist orthopedists to improve the efficiency and accuracy of calcaneus fracture assessment.

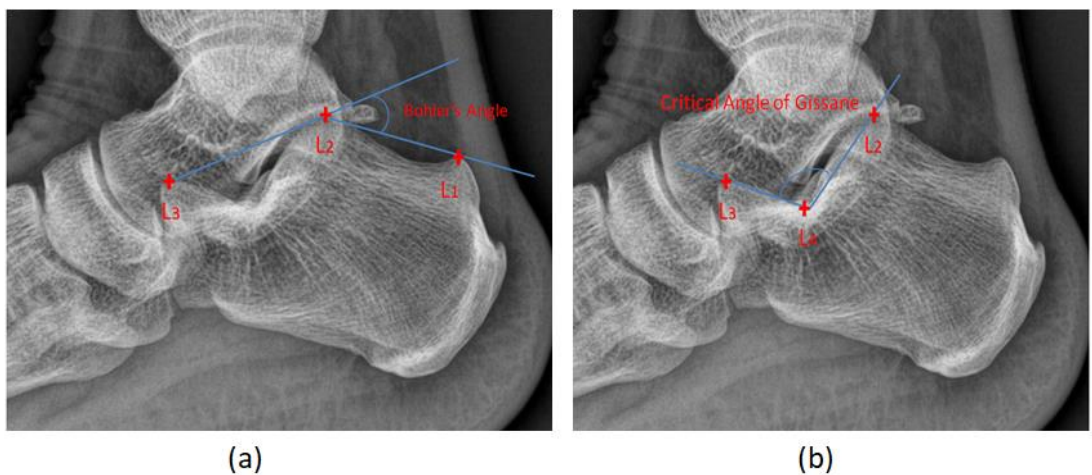

Figure 1 BA and CAG in lateral calcaneus radiograph. (a) Bohler's Angle. (b) Critical Angle of Gissane.

The accurate localization of landmarks is the key to the measurement of calcaneal angles. The localization of anatomical landmarks is an important and challenging step in the clinical workflow for therapy planning and intervention. Classifier-based anatomical landmark localization classifies the positions of selected interest points [10]. To achieve high accuracy, the candidate points have to be dense for purely classifier-based approaches. Regression-voting-based approaches, as alternatives, have become the mainstream [11][12][13][14]. Approaches that use local image information and regression-based techniques can provide much more detailed information. Work on Hough Forests [15] has shown that objects can be effectively located by pooling votes from Random Forest (RF) [16] regressors.

Currently, deep learning algorithms, in particular convolutional neural networks [17], have rapidly become a methodology of choice for analyzing medical images. Since the profound performance improvement on ImageNet challenge [18] contributed by AlexNet [19], many works have achieved powerful performance on a variety of tasks including classification, segmentation, and detection. Some recent methods using CNNs in anatomical landmark localization have shown promising results [20][21], where heatmaps are used to locate landmarks at certain positions. There are also many attempts to use CNN to detect bone fractures [22][23]. Previous computer-assisted diagnosis of calcaneus fracture concentrates on calcaneus CT [24][25]. For calcaneus radiograph, it is challenging for fracture identification because of the inconsistent rotation and view, the complex anatomy, and the lack of medical information comparing to CT.

Some work [26] indicates that rotation invariance might not be necessary for most medical contexts; however, the lateral foot radiographs are not among them due to fickle variance of rotation due to the different postures of patient and condition of X-ray machine as shown in Figure 2. Most landmark detection method based on regression is not rotationally invariant as opposed to classifier-based method, because regressive displacement in image coordinate is not invariant to rotation of objects.

When implementing a regression-voting-based method in datasets with bigger rotations, it is suggested to initially using a 2D registration algorithm such as [27] to roughly align the images among each other before analysis [28]. CNN models are empirically regarded to be invariant to a moderate translation of input image. Meanwhile, it is also known that most conventional CNN models are not explicitly rotation-invariant [29]. It's a common practice to augment images by spatial and intensity transform during training, which enables the model to become rotation-invariant implicitly. But even if the model perfectly learns the invariance on the training set, there is no guarantee that this will generalize [30]. In addition, some works succeeded to explicitly encode this characteristic in the model [29][31][32]. However, these methods are either only applied to classification task, or hard to combine with landmark localization CNN structure.

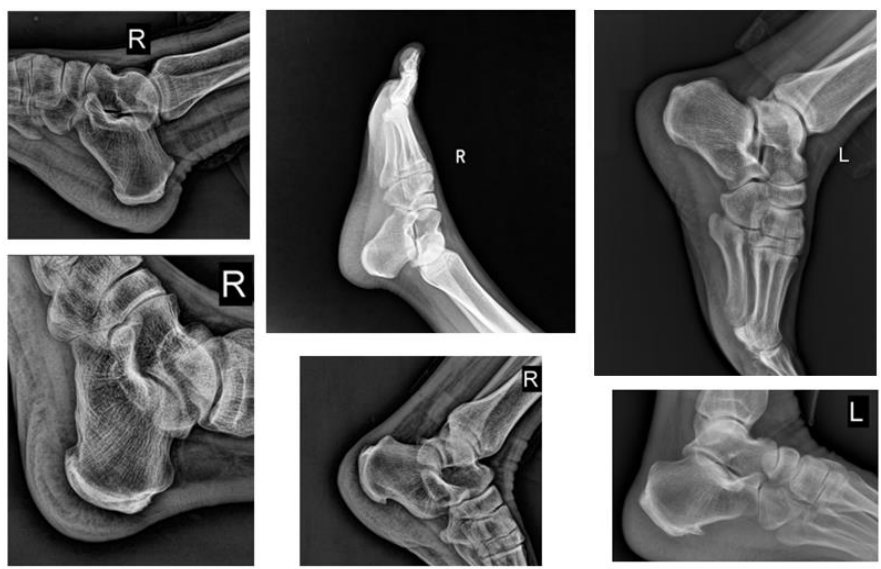

Figure 2 Different rotation and view in lateral foot radiograph

This study proposed a new analysis system of calcaneus fracture detection in radiographs, including landmark detection, measurement of BA and CAG, fracture identification, and fracture region segmentation. Firstly, for landmark detection, we proposed a coarse-to-fine Rotation-Invariant Regression-Voting (RIRV) method based on regressive Multi-Layer Perceptron (MLP) [33] and Scale Invariant Feature Transform (SIFT) [34] patch descriptor. Secondly, the locations of anatomic landmarks are used to calculate BA and CAG which can be used by orthopedists to assess the condition of fracture. Thirdly, for fracture identification and segmentation, we designed a multi-task convolution neural network (CNN) named U-Net [35] with auxiliary classification output (U-Net-CH). The flow chart of the proposed analysis system is shown in Figure 3.

The contribution of the study can be summarized as: (1) We proposed a regression-voting-based landmark detection method RIRV. The method normalizes the regressive displacement by the scale and orientation of each SIFT patch, which results in the explicit scale-rotation invariance without any need of data augmentation or rough alignment; (2) The RIRV implements a novel screening method, half-path double voting (HDPV), to remove unconfident voting candidates caused by far, noisy or blurry image patches; (3) We implemented U-Net with auxiliary classification head (U-Net-CH), which makes use of the features extracted by encoder and decoder, to generate fracture judgment in addition to fracture region segmentation; (4) The landmarks' location is used as prior knowledge. The input ROI of the U-Net-CH network is normalized by the result of detected landmarks to uniform view, orientation, and scale so that satisfactory performance can be obtained with a small number of training data.

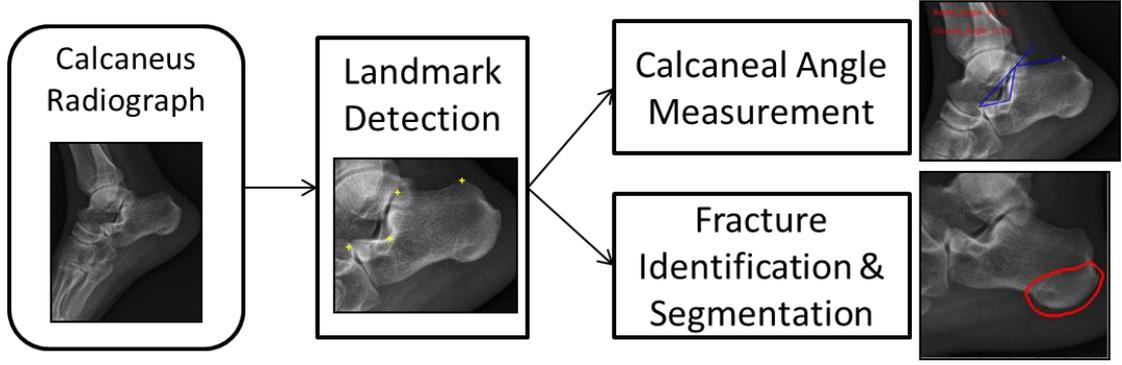

Figure 3 Flow chart of the calcaneus radiograph analysis system.

## 2. Methodology

### 2.1. Landmark Detection: Rotation-Invariant Regression-Voting

#### 2.1.1. Restrictions of Conventional Regression-Voting and Its Rotation-Invariant Improvement

In conventional regression-voting approaches, a regressor, such as Support Vector Regression (SVR) [36] or regressive MLP[33], is trained for each landmark. The image area that an image descriptor (like SIFT [34] or SURF [37]) locates and spans is called as an image patch $p$. For each landmark $L_i(x, y)$, $N$ image patch features $f_j(p(x, y, s, \theta)), j = 1,2 \ldots N$ are extracted from a set of random image patches $p_j(x, y, s, \theta)$ sampled in the sampling region (SR), where $x$, $y$, $s$ and $\theta$ are x coordinate of patch center, y coordinate of patch center, scale and orientation of the image patch, respectively. Next, the set of displacements $d_j(x, y)$, from the center of random patches to the landmark ground truth, are calculated. Then, a regressor $\delta = \mathrm{Tr}(f_j, d_j)$ is trained to predict the relative displacements $\bar{d}_j$ from $p_j$ to $L_i$.

The displacements can be used to predict most likely position of the landmark based on voting (mean or highest density) of candidate $c_j = p_j + \bar{d}_j$. The regressive displacement is illustrated in Figure 4.

However, though image patch descriptors such as the SIFT feature, are extracted with scale and orientation, the displacement is on the base of image coordinate: *x* represents pixel column and *y* represents pixel row, which is variant to the rotation of calcaneus. The regressive displacements learned from Figure 4(a) are messed when performing on a rotated object and a scaled object, as illustrated in Figure 4(b) and Figure 4(c). Though in most medical applications when the rotation and scale of objects are controlled to a fine extent and the training set can be artificially augmented by randomly rotating, data augmentation can hardly deal with calcaneal radiograph in such dominant rotation variance. In this study, SIFT feature descriptors are used to represent image patches as shown in Figure 5.

Therefore, we proposed a rotation-invariant regression voting (RIRV) landmark detection method to normalize the regressive displacement $d_j$ which bases on the coordinate of the whole image, to $d^{norm}_j(x^{norm}, y^{norm})$ which bases on the coordinate defined by scale and orientation of the corresponding $p_j$. This method results in the explicit scale-rotation invariance without the need for image augmentation.

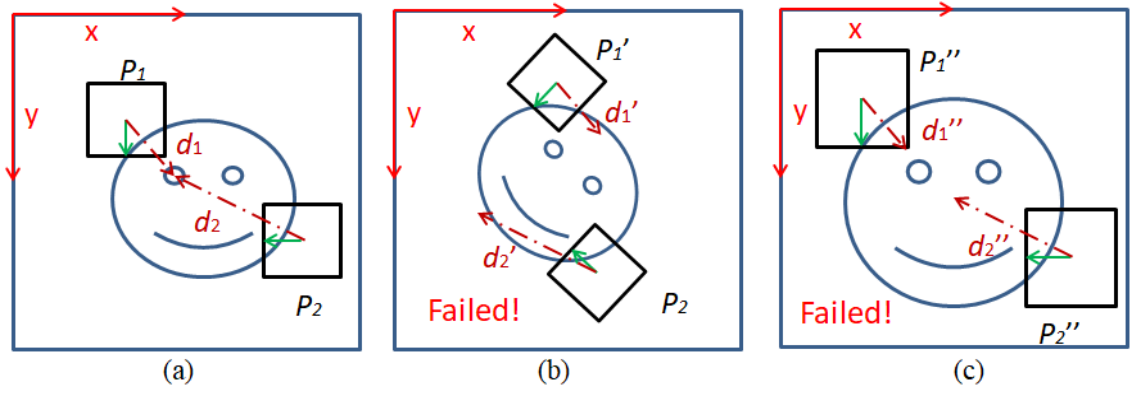

Figure 4 Regressive displacement relationship. Black squares represent image patches. Green arrows are the orientations of the patches. Crimson dash lines are the displacement from the center of patches to the target left eye. (a) Regressive displacemenst learned in original image. (b) The regressive displacements in subfigure (a) performing on rotated image. (c) The regressive displacemenst in subfigure (a) performing on scaled image.

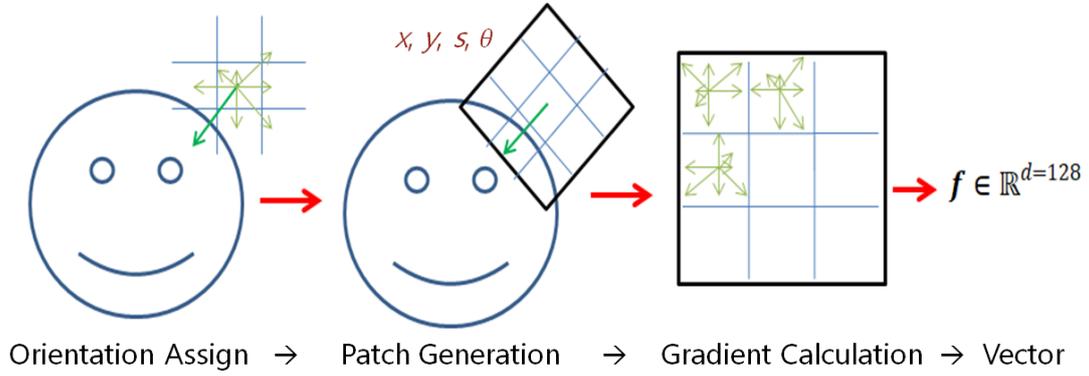

Figure 5 SIFT feature extraction. An image patch $p(x, y, s, \theta)$ is represented as $f \in \mathbb{R}^{d=128}$

### 2.1.3. Displacement Normalization and Denormalization

Displacement normalization is the key to Rotation-Invariant Regression-Voting. The original displacements $d_j(x, y)$ start from the center of random patches to the ground truth of the landmark. The normalized displacement $d^{norm}_j(x^{norm}, y^{norm})$ is calculated as:

$$d^{norm}_j(x^{norm}, y^{norm}) = \begin{bmatrix} x^{norm} \\ y^{norm} \end{bmatrix} = \frac{\text{rotate}(d_j(x,y), \theta_j)}{s_j}, \quad (\text{Eq. 1})$$

$$\text{rotate}(d(x,y), \theta) = \begin{bmatrix} \cos(\theta) & -\sin(\theta) \\ \sin(\theta) & \cos(\theta) \end{bmatrix} * \begin{bmatrix} x \\ y \end{bmatrix}, \quad (\text{Eq. 2})$$

where *rotate*( • ) means rotating a vector by $\theta$ degree. After normalization, $d^{norm}_j$ can be seen in the coordinate of the corresponding $p_j$, which is relative to both orientation $\theta_j$ and scale $s_j$.

Denormalization from $d^{norm}_j$ to $d_j$ can be defined as:

$$d_j(x,y) = \begin{bmatrix} x^{norm} \\ y^{norm} \end{bmatrix} = \text{rotate}\big(d^{norm}_j(x^{norm}, y^{norm}), -\theta_j\big) * s_j, \qquad (Eq.\ 3)$$

where $d_j(x,y)$ is in the original image coordinate.

The normalization in radiographs can be illustrated in Figure 6, where red arrows and green arrows represent image coordinate and patch coordinate, respectively. Though the displacements $d_j$ in image coordinates are different in two images, the displacements $d^{norm}_j$ in its corresponding patch coordinate are the same.

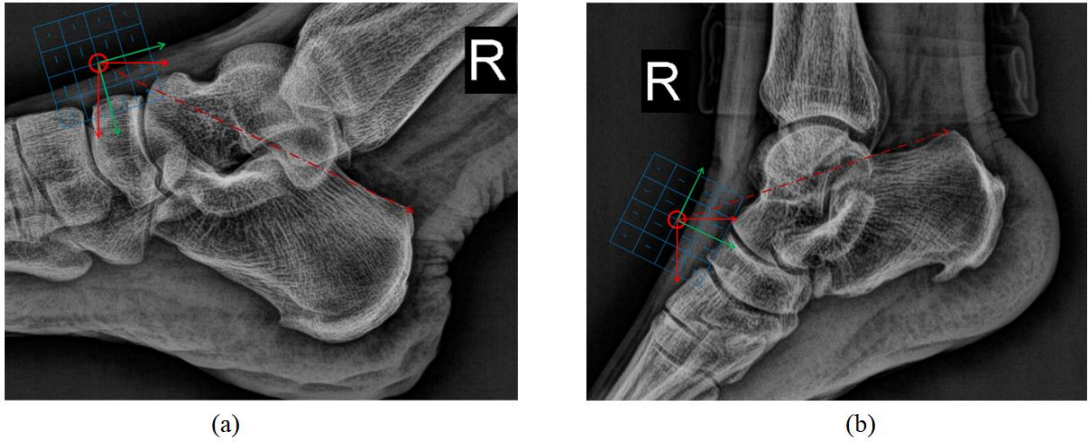

(a)           (b)

Figure 6 Displacement normalization in calcaneus radiograph. (a) Radiograph 1. (b) Radiograph 2. Green and red arrows are the patch-based coordinates and image coordinates, respectively. Crimson dash lines are the displacements from the centers of patches to $L_1$.

### 2.1.4. Procedures of RIRV

RIRV method adopts a multi-stage coarse-to-fine strategy with four stages. We implement regressive MLP [33] with two-cells output (corresponding to *x* and *y*) as the displacement predictor. The architecture is shown in Figure 7. In each stage, for each landmark $L_i(x, y), i = 1,2,3,4$, a regressive MLP $\boldsymbol{\delta}_{h,i}$ is trained by the set of $N$ pairs $\{(\boldsymbol{d}^{norm}_j, \boldsymbol{f}_j)\}$ (displacement and SIFT feature) sampled randomly in the sampling region (SR), where $h$ is the number of the stage. For the first stage, SR is the whole radiograph. For the other stages, the SR is the neighborhood of each landmark. The width of each SR is $D_{SR;h}$ pixels, where $h$ is the number of stage. The orientation $\theta$ of SIFT patch is automatically assigned by the dominant gradient angle in the patch according to SIFT algorithm in the first and second stages. In the third and fourth stages, $\theta$ is the slope angle of line $L_1 - L_3$ obtained from the previous stage plus a random perturbing in range $[\Delta\theta_{min;h}, \Delta\theta_{max;h}]$ to make use of the rotation information of the coarse detection. In the third and fourth stages, all radiographs are horizontally flipped to toe-left based on landmarks' location. Toe-left or toe-right can be determined by the relative location of landmarks. For toe-left calcaneus, $L_2$ and $L_4$ are in the clockwise and counter-clockwise direction of vector $L_1 - L_3$ respectively, and vice versa. This conversion constrains stage three and four to only toe-left radiographs; meanwhile it restricts the feature domain to improve accuracy. The scale of the patch is randomly and flatly assigned in range in $[s_{min;h}, s_{max;h}]$. Intuitively, a large sampling region would benefit the robustness to inaccuracy of previous stages; meanwhile, the regressors would be hard to train and prediction will be slowed if the sampling region is too large. A higher range of $[\Delta\theta_{min;h}, \Delta\theta_{max;h}]$ and $[s_{min;h}, s_{max;h}]$ contributes to scale and rotation invariance. A wider sampling region $D_{SR;h}$ contributes to the tolerance to the inaccuracy of the previous stage.

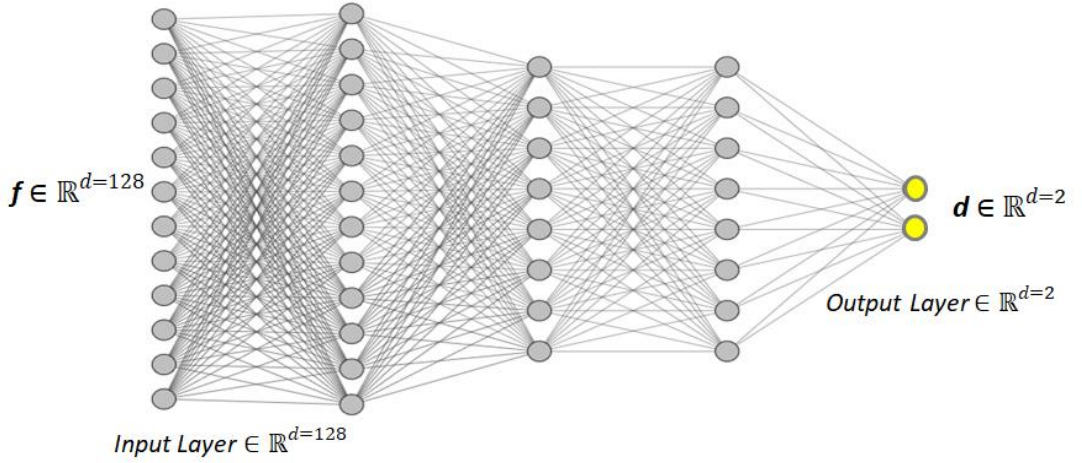

Figure 7 Regressive multi-layer perceptron with 3 hidden layers. The sizes of the layers are 128, 200, 100, 100, and 2 from input to output layer. Each layer is flowed by a batch normalization layer [38] and a Tanh activation, except the output layer.

The prediction of RIRV is the proper reverse of training. Latter stages are designed to refine the results of previous stages. In each stage, for each landmark $L_i(x,y), i=1,2,3,4$, the MLP $\boldsymbol{\delta}_{h,i}$ can predict $\overline{\boldsymbol{d}_j^{norm}} = \boldsymbol{\delta}_{h,i}(\boldsymbol{f}_j)$ by $N$ $\boldsymbol{f}_j$ sampled randomly in the SR. SR is the whole radiograph for $h=1$ or neighborhood of the predicted location of the previous stage for $h=2,3,4$. Then, $\overline{\boldsymbol{d}_j^{norm}}$ is denormalized to $\overline{\boldsymbol{d}}_j$. The voting candidates can be calculated as $\boldsymbol{c}_j = \boldsymbol{p}_j + \overline{\boldsymbol{d}}_j$. The image patch parameters $[\Delta\theta_{min;h}, \Delta\theta_{max;h}]$, $[s_{min;h}, s_{max;h}]$ and $D_{SR;h}$ should ensure that the image patches in prediction are less variant than in training (e.g. the scale range in prediction should be smaller than training). After stage two, all radiographs are horizontally flipped to toe-left based on detected landmarks' location from the stage two.

In the prediction, a screening method named Half-Path Double Voting (HPDV) is proposed to remove unconfident voting candidates caused by far, noisy, or blurry image patches. For each image patch $p_j$ in the first three prediction stage, a half-path patch $p_j^{half}(x^{half}, y^{half}, s, \theta)$ is sampled at the midpoint between voting candidate $c_j$ and the center of $p_j$. The location of $p_j^{half}$ is calculated as $[x^{half}; y^{half}] = p_j + \bar{d}_j/2$. The half-path patch will operate second voting by the same method to calculate $c_j^{half}$. The result of $c_j^{half}$ is valid as final voting candidates $c_j^{valid}$ of the stage as:

$$c^{valid} = \{c^{half} \mid \|c_j^{half} - c_j\| < Th_h\}, \quad \text{(Eq. 4)}$$

where $Th_h$ is the pixel deviation threshold empirically assigned to each stage. The illustration of HPDV is shown in Figure 8. This method can be seen as a screening method to validate whether each voting candidate is credible. An example of HDPV is illustrated in Figure 9, where $c$ is denoted as green point and $c^{valid}$ is denoted as red point. The possibility density maps obtained by Kernel Density Estimation (KDE) of $c$ and $c^{valid}$ are illustrated by Figure 9(b) and Figure 9(c), respectively. It can be seen that $c^{valid}$ is more concentrated and accurate. The voted landmarks location $\bar{L}_\iota$ is located at the highest possibility in the density map.

The flowchart of the RIRV landmark detection method is shown in Figure 10. An example of the result of each stage in the RIRV algorithm is shown in Figure 15 in the Appendix.

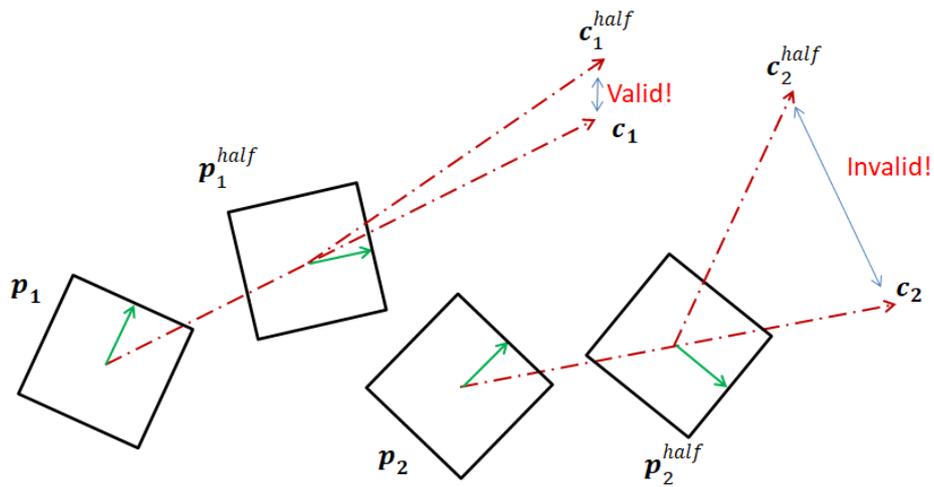

Figure 8 Illustration of Half Path Double Voting. Black squares represent image patches and green arrows are the orientations of the patches. Crimson dash lines represent the predicted displacements from the center of patches to a target landmark. If the two votes' distance is further than a threshold, the votes will be discarded.

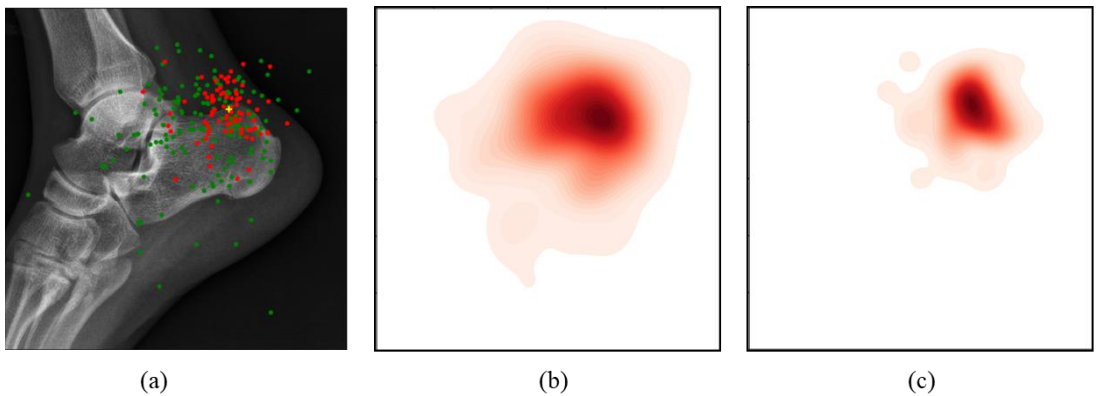

(a)  (b)  (c)

Figure 9 Half-Path Double Voting in stage one of RIRV. (a) The voting result on a radiograph. Original voting candidates $\mathbf{c}$ are denoted as green point and screened valid candidates $c^{valid}$ are denoted as red point. (b) The heat map of $\mathbf{c}$. (c) The heat map of $c^{valid}$. It can be seen that $c^{valid}$ is more concentrated and accurate.

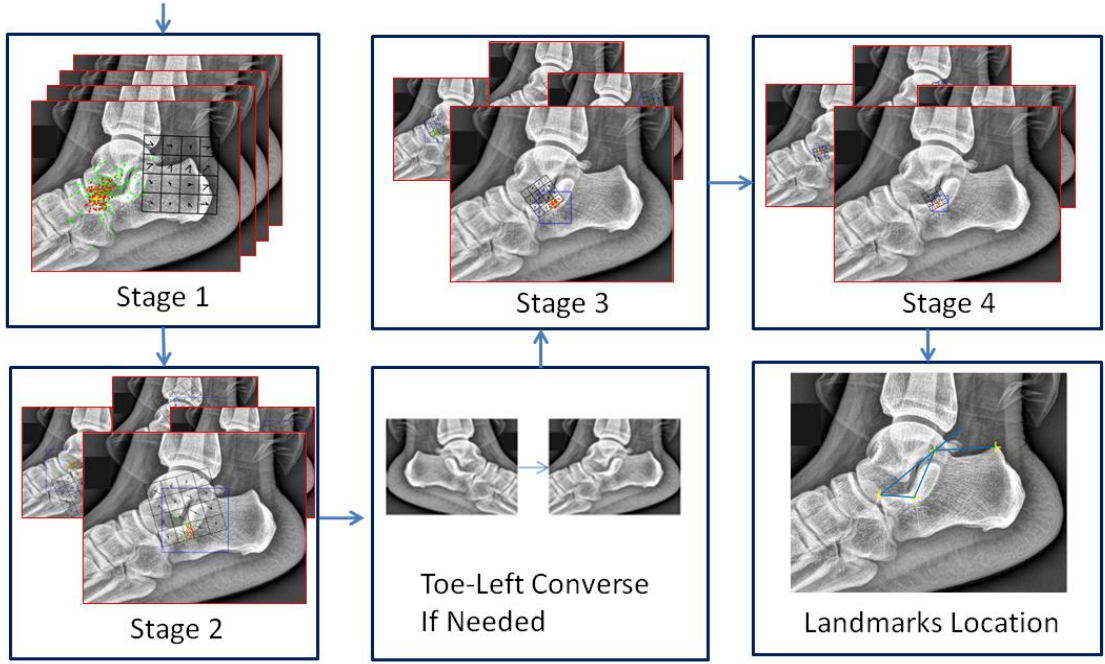

Figure 10 Flow chart of landmark detection in lateral calcaneus radiograph based on RIRV

## 2.2. Calcaneal Angle Calculation

After four landmarks are located, the anatomical angles can be measured. The Bohler's Angle is calculated by:

$$\angle \text{BA} = 180° - \angle L_1L_2L_3 = \frac{L_2 - L_3 \cdot L_1 - L_2}{\|L_2 - L_3\| * \|L_1 - L_2\|} * \text{sign}(A \times B) * \frac{180}{\pi}, \quad (\text{Eq. 5})$$

where $\angle L_1L_2L_3$ is the angle beneath the vertex $L_2$; $sign(\cdot)$ gives the sign in the parentheses, which means $\angle L_1L_2L_3$ can also be a reflex angle and $\angle$BA can be negative. The Critical Angle Gissane is calculated by:

$$\angle \text{CAG} = \angle L_2L_4L_3 = \frac{(L_2 - L_4) \cdot (L_3 - L_4)}{\|L_2 - L_4,\| * \|L_3 - L_4\|} * \frac{180}{\pi}, \quad (\text{Eq. 6})$$

## 2.3. Fracture Identification and Fracture Region Segmentation

### 2.3.1. Input ROI Normalization by Detected Landmarks

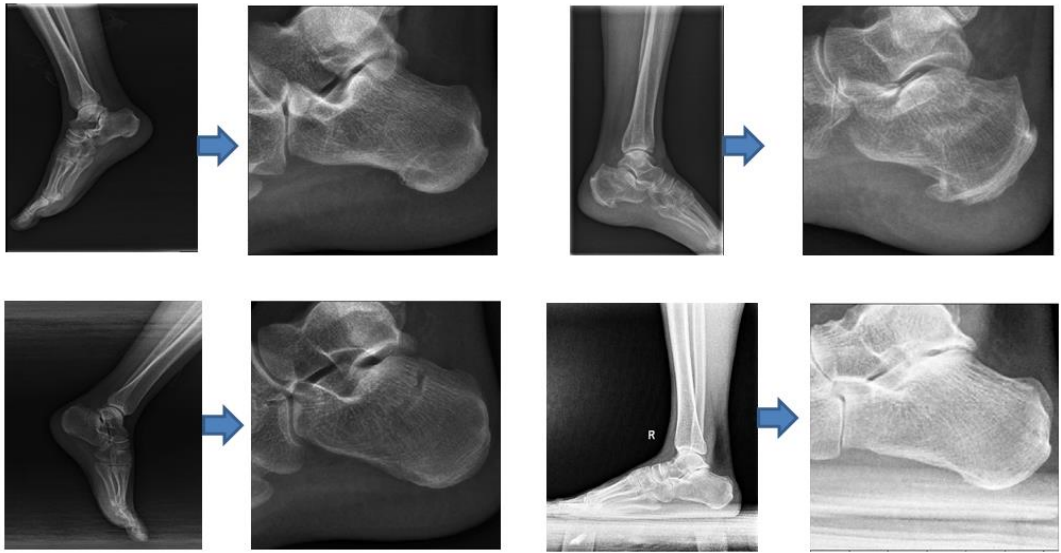

Figure 11 Input ROI normalization.

The original calcaneus radiographs are usually subjected to larger fickle view that includes the whole foot or even lower calf. Therefore, a preprocessing and normalization procedure is necessary before feeding the image into CNN. The landmarks' location detected by RIRV is used as medical prior knowledge. The input ROIs are normalized by the result of detected landmarks to uniform view, orientation, and scale so that satisfactory performance can be obtained with a small number of training data. To be specific, the radiograph is converted to toe-left and rotated to $L_1$-$L_3$ horizontal. ROI of square shape is cropped and resized to 320×320 pixels based on the location of the landmarks to contain the whole calcaneus. The ROI is then contrast-enhanced by the CLAHE algorithm [39]. Some examples of ROI cropping and normalization are shown in Figure 11.

**2.3.2. U-Net with Auxiliary Image Classification Head**

In 2015, Ronneberger et al. [35] proposed an architecture (U-Net) consisting of a contracting path to capture context and a symmetric expanding path that enables precise segmentation. In this section, an auxiliary classification output is added to the U-Net network to generate judgment of fracture in addition to fracture region segmentation; therefore, the network makes fracture identification and segmentation as multi-task. The proposed structure is called U-Net-CH (classification head). The encoder, i.e. backbone, of U-Net-CH adopts pre-trained networks. The number of input channels for the first convolution layer of the network (stem layer, usually a 7×7 convolution followed by a BN [38] and a ReLU) is converted to one because radiographs are gray-scale images. The final linear classification layer of the backbone is deprecated. The feature maps generated by the encoder are used in the decoder, i.e. the expansive path. Similar to the original U-Net, each decoder layer begins with an up-sampling layer to multiply the feature maps spatial size by two. The up-sampling is followed by encoder-decoder features concatenation and a residual block. The final segmentation head is a 3×3 convolution followed by a sigmoid activation to convert feature maps to segmentation output.

The auxiliary classification head makes use of the features from both encoder and decoder. The final feature maps of the encoder are fed into a residual block first. The final feature maps of the decoder are adaptively max pooled to the same spatial size of those of the encoder, and fed to two residual blocks. Then, the encoder and decoder features are concatenated and followed by a residual block, a global max pooling, and a binary classification output with Softmax activation. The whole architecture of U-Net-CH is shown in Figure 12. The classification head can synthesize the feature maps used for segmentation to help image classification

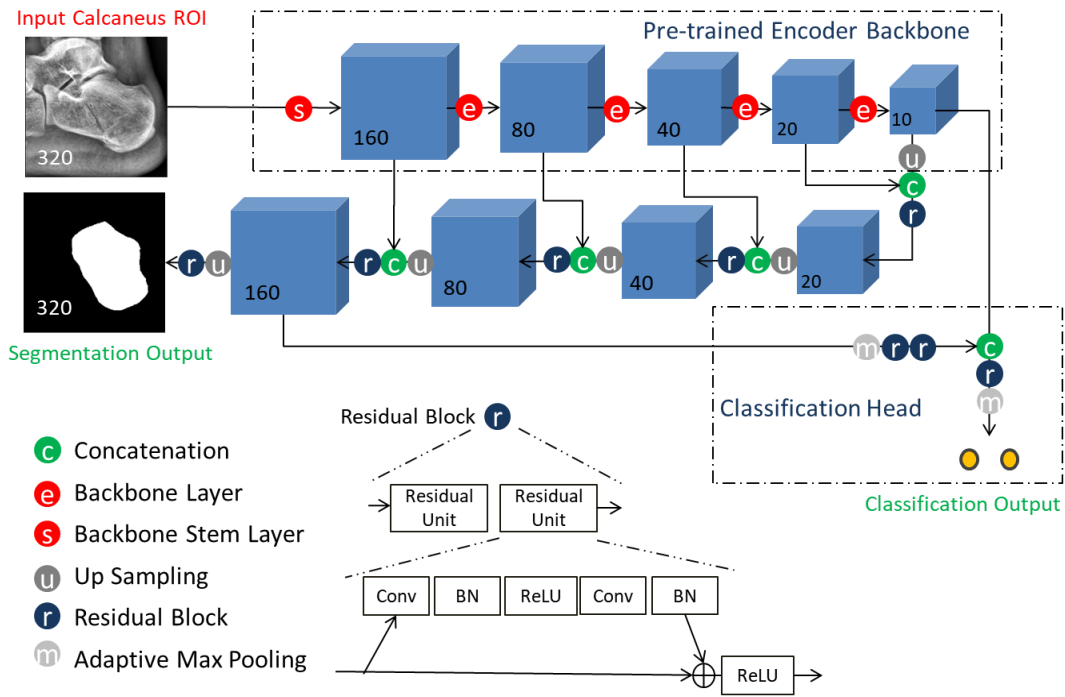

Figure 12 U-Net with auxiliary classification head and Pre-trained Encoder.

# 3. Experiments and Results

## 3.1. Datasets

Evaluation of the proposed system performs on the calcaneus radiograph dataset collected from the First Affiliated Hospital of Jinzhou Medical University. Four landmarks were manually annotated by experienced orthopedists and judgments of fracture were made (normal, intra-articular fractured, or extra-articular fractured). In fractured radiographs, the regions of fracture were annotated as polygons. The dataset includes 1254 radiographs of normal calcaneus, 957 radiographs of fractured calcaneus (757 intra-articular and 200 extra-articular). Among the fractured radiographs, 69 are annotated as suspicious fractured. However, all suspicious radiographs are regarded as fractured in this study as they should be reported by the analysis system for further clinic evaluation. The image resolutions vary from 536×556 to 2969×2184 pixels.

In evaluation of RIRV, one-fourth of the radiographs are separated as the test set and the others for the training set. In evaluation of U-Net-CH, we adopt 4-fold cross validation. There is no pixel-mm ratio annotated in radiographs, so this study set the distance between $L_1$ and $L_3$ as a referencing length to evaluate detection performance in millimeter. It is assumed that the distance between $L_1$ and $L_3$ as 70mm and there is no prominent difference across individuals.

## 3.2. Evaluation of RIRV Landmark Detection

The image patch parameters of RIRV in training and prediction are set as shown in Table 5 and Table 6 in the Appendix. The values of parameters ensure that the image patches in prediction are less versatile and fickle than in training. SIFT features are extracted by cyvlfeat 0.4.6 (a Python wrapper of the VLFeat library [40]). The MLPs are implemented by the PyTorch 1.6.0 framework and trained for 100 epochs with an initial learning rate of 0.05 and a drop factor of 0.5 every 30 epochs.

The first evaluation criterion is Mean Radius Error (MRE) associated with Standard Deviation (SD) of the four landmarks and Mean Angle Error (MAE) of two anatomical angles. The Radial Error (RE) is the distance between the predicted position and the annotated true position of each landmark. Another criterion is the Success Detection Rate (SDR) with respect to 2mm, 4mm, and 6mm. If RE is less than a precision range, the detection is considered as a successful detection in the precision range; otherwise, it is considered a failed detection. The SDR with Real Length Error less than error precision p is defined as:

$$SDR\_p = \frac{\{RE_i < p\}}{N} \times 100\%, 1 \leq i \leq N, \quad (Eq. 7)$$

where $p$ denotes the three precision range: 2mm, 4mm, and 6mm

Experimental results of calcaneal landmarks detection are shown in Table 1. The average MRE and SD of four landmarks in all radiographs are 1.85 and 1.61mm, respectively. The results of calcaneal angle measurements are shown in Table 2. The detection accuracy of landmarks and angels varies from one to another; $L_1$ shows good accuracy while $L_2$ shows the worst. There are three main reasons. First, sometimes $L_2$ and $L_4$ are not imaged clearly because they are blocked by tarsus.

Second, the definition of $L_2$, i.e. the apex of the posterior facet of the calcaneus, is hard to precisely carry out in annotating because the posterior facet is an arc and its apex is widely ranged. Third, in fractured calcaneus radiograph, the anatomical structure of landmarks, especially for $L_2$ and $L_3$, is changed by fracture of the calcaneus, and the feature varies from one fractured instance to another. Some examples of imprecise localization are shown in Figure 13.

Comparison between proposed RIRV with Multiresolution Decision Tree Regression Voting (MDTRV) [12] and SpatialConfiguration-Net (SCN) [21] is shown in Table 3. The MDTRV is a regression-voting-based method tested on the cephalogram database of the 2015 ISBI Challenge. SCN is a CNN-based heatmap regression method for anatomical landmarks localization. Furthermore, the proposed method is evaluated on an augmented test set, in which all radiographs were randomly rotated (up to 360°) to prove the explicit rotation-invariance character of RIRV. The result is shown in Table 3 headed with RIRV-RTS (rotated test set).

The result shows that the proposed method outperformed MDTRV and SCN. The high SD of MDTRV is mainly caused by the complete misdetection of some radiographs with great rotation that seldom appears in the training set. The CNN-based method is able to constrain the prediction near the calcaneus for radiographs with great rotation, but there is still a lot of room to improve. In addition, there is no prominent difference in the result of RIRV performing on the original and rotated test set.

Table 1 Statistical results of landmark detection

|  | MRE (mm) | SD (mm) | SDR_2mm (%) | SDR_4mm (%) | SDR_6mm (%) |
|---|---|---|---|---|---|
| $L_1$ | 1.05 | 0.90 | 88.84 | 98.76 | 99.59 |
| $L_2$ | 2.73 | 2.22 | 47.11 | 78.51 | 92.15 |
| $L_3$ | 1.73 | 1.63 | 68.18 | 95.87 | 97.93 |
| $L_4$ | 1.92 | 1.69 | 64.88 | 92.98 | 96.28 |
| Average | 1.85 | 1.61 | 67.25 | 91.53 | 96.49 |

Table 2 Statistical results of calcaneal angles measurements

|  | MAE(°) | SD(°) |
|---|---|---|
| BA | 3.84 | 4.20 |
| CAG | 6.19 | 6.23 |

Table 3 Statistical Comparison between RIRV with MRDTV and SCN. RIRV is also tested on a random rotated test set, headed with RIRV-RTS

| Methods | MRE (mm) | SD (mm) | SDR_2mm (%) | SDR_4mm (%) | SDR_6mm (%) |
|---|---|---|---|---|---|
| MDTRV | 3.25 | 11.07 | 62.31 | 86.63 | 90.15 |
| SCN | 3.32 | 3.24 | 59.13 | 72.25 | 90.05 |
| RIRV | 1.85 | 1.61 | 67.25 | 91.53 | 96.49 |
| RIRV-RTS | 1.87 | 1.72 | 68.88 | 91.32 | 97.01 |

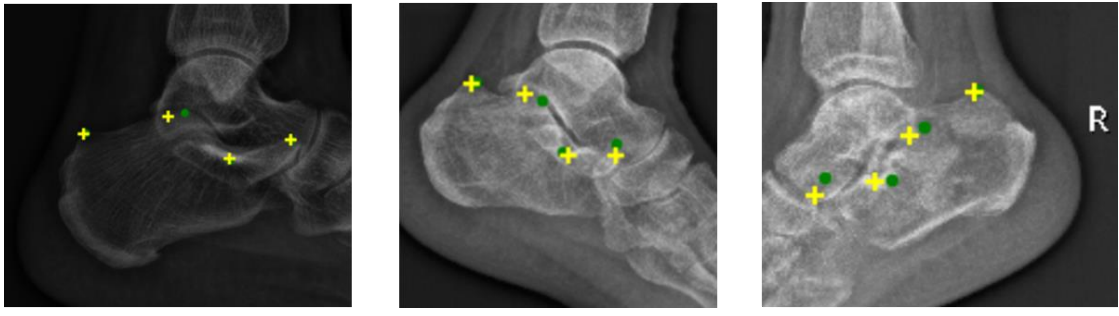

Figure 13 Examples of imprecise landmark localization. Green dots represent ground truth landmarks. Yellow crosses represent predicted landmarks.

### 3.3. Evaluation of Fracture Identification and Segmentation

The input images are cropped and normalized from radiographs according to the landmark locations, as described in section 2.3.1. All ROIs are resized to 320×320 pixels before feeding into networks. The training labels of the ROIs are the annotated labels of the corresponding radiographs: normal or fractured. During training, images are augmented with random rotation between -30° and +30°, random translation within 30 pixels, random contrast, and random brightness; there is no random flip. The augmentation is limited to a low extent to maintain the normalized information resulted from multi-region image cropping.

For classification evaluation, F1-Score, Recall, and Precision are defined as:

$$\text{Recall} = \frac{TP}{TP + FN}, \quad (\text{Eq. 8})$$

$$\text{Precision} = \frac{TP}{TP + FP}, \quad (\text{Eq. 9})$$

$$\text{F1 Score} = 2\frac{\text{Recall} \times \text{Precision}}{\text{Recall} + \text{Precision}}, \quad (\text{Eq. 10})$$

where TP, TN, FN are the number of correctly predicted fractured radiographs, correctly predicted normal radiographs, and fractured radiographs wrongly predicted as normal, respectively. For segmentation evaluation, IoU score (intersection over

union) is defined as:

$$IoU = \frac{|X \cap Y|}{|X| + |Y| - |X \cap Y|}, \quad \text{(Eq. 11)}$$

where *X* and *Y* are the predicted and ground truth segmentation, respectively.

The proposed U-Net-CH is implemented with Effcientnet-b3 [41] backbone and compared with several mainstream classification CNNs including Densenet169 [42], Inception-v4 [43] ResNeXt50_32x4d [44] and Effcientnet-b3 [41] and vanilla segmentation U-Net [35] with Effcientnet-b3 backbone. The networks are built and run on PyTorch 1.6.0 framework and pre-trained on ImageNet [18]. The number of input channels for the first convolution layer of the networks is converted to one because radiographs are gray-scale images. The linear classification layers of the classification networks are discarded and replaced by a global max-pooling layer which is followed by linear classification output. All networks are trained for 60 epochs with a global initial learning rate of 1e-4 and a drop factor of 0.3 every 20 epochs. Adam [45] optimizer is adopted with a decay factor of 0.999, momentum of 0.9, and epsilon of 1e-8. The optimization loss function for classification and segmentation is Cross-Entropy Loss and Dice Loss respectively. The input ROIs of the proposed U-Net-CH as well as all baseline networks are normalized as described in section 2.3.1. In addition, to prove the effectiveness of our normalization method, we trained and evaluated U-Net-CH without ROI normalization for comparison, in which the ROIs are only cropped according to landmark locations but without normalization of view, orientation, and scale.

Table 4 Statistical comparison of fracture identification and segmentation.

| Networks | Seg. IoU | Cls. F1-Score(%) | Cls. Precision(%) | Cls. Recall(%) Overall | Cls. Recall(%) Intra | Cls. Recall(%) Extra |
|---|---|---|---|---|---|---|
| Densenet169 | - | 95.72 | 97.74 | 93.82 | 97.31 | 80.66 |
| ResNext50 | - | 95.72 | 97.74 | 93.84 | 97.74 | 79.34 |
| Inception-V4 | - | 95.74 | 97.61 | 93.94 | 97.86 | 78.95 |
| Efficient-b3 | - | 95.96 | 97.24 | 94.77 | 98.26 | 81.86 |
| U-Net-CH | **0.5859** | **96.55** | **98.17** | 94.99 | 98.13 | 83.14 |
| U-Net | 0.5834 | 96.21 | 96.84 | **95.60** | **98.25** | **85.64** |
| U-Net-CH w/o N | 0.5674 | 94.45 | 98.06 | 91.11 | 97.08 | 68.93 |

The statistical results are shown in Table 4, showing that U-Net-CH outperforms all baseline classification networks. In the table, IoU is given for segmentation. F1-score, precision, and recall are given for fracture identification. All networks except 'U-Net-CH w/o N' are performed with ROI normalization. We take the highest value in the segmentation map as predicted fracture probability for the vanilla U-Net as it is a pure segmentation model. As a result, the recall of vanilla U-Net can be higher at the cost of lower precision. The result of U-Net-CH without ROI normalization is the worst, which proves the efficiency of our normalization method. The sensitivity of extra-articular fracture is lower than intra-articular because of its trivial feature. The segmentation of U-Net-CH is more sensitive to trivial extra-articular fractured areas than that of vanilla U-Net as shown in Figure 14. The original labeling of fracture is polygon-shaped and quite rough, including the region that is not fractured and not even calcaneus, as shown in Figure 16 in the Appendix;

therefore, IoU computed with the label mask cannot fully represent the performance in this case. Some examples of the whole calcaneus analysis system including landmark detection, BA and CAG measurement, and fracture region segmentation are shown in Figure 17 in the Appendix.

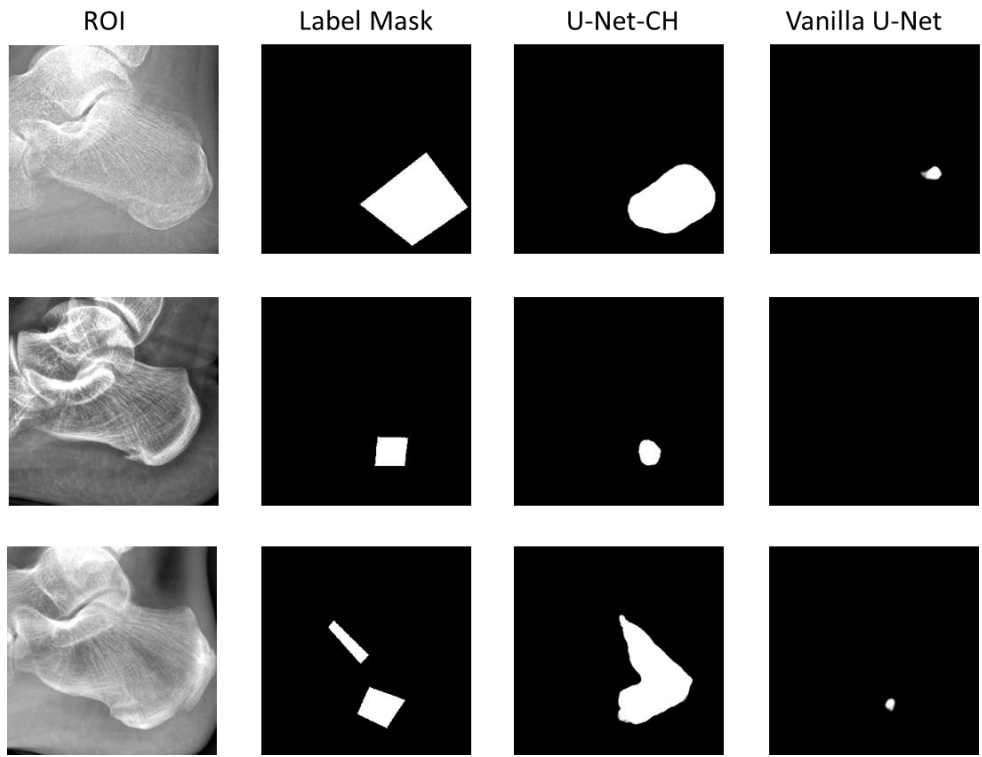

Figure 14 Examples of fracture segmentation by U-Net-CH and vanilla U-Net.

## 4. Discussion

The result shows that a powerful calcaneus radiograph analysis system including anatomical angles measurement, fracture identification, and fracture region segmentation have been built. The proposed RIRV landmark detection method is suitable for calcaneal landmark detection for its explicit rotation-invariant characteristic. The approach can deal with 360° rotation of images without the need for training augmentation. The experimental result shows that RIRV can achieve MRE of 1.85mm and SD of 1.61mm; the MAE of BA and CAG is 3.84 and 6.19 respectively. In addition, a subset of 20 in the dataset is randomly selected and two experienced orthopedists are asked to manually annotate the four landmarks to calculate inter-observer variability. The inter-observer MRE and SD between two observers are 1.53mm and 1.39mm, respectively. The inter-observer MAE of BA and CAG is 2.21° and 4.95°, which means that the performance of RIRV is a little lower than manual annotation but still satisfactory. The accuracy of CAG is lower than BA; however, the analysis as well as literature [46] shows that inter-observer variability for CAG is poor as well. Though the statistical shape model is needless in the proposed method due to the small number of landmarks, RIRV can be easily integrated with Active Shape Models (ASMs) [47] or Constrained Local Models (CLMs) [48].

The proposed CNN structure U-Net-CH, i.e. U-Net with classification head, can help to screen fractured calcaneus and detect fracture region simultaneously. Due to the ROI normalization by the location of landmarks, the U-Net-CH could deal with the subtle feature of different fractures, presenting good performance with classification F1-score of 96.55% and recall of 94.99%, segmentation IoU-score of 0.586. The sensitivity of extra-articular fracture is lower than intra-articular because of its trivial feature. The segmentation of U-Net-CH is more sensitive to trivial extra-articular fractured areas than that of vanilla U-Net. The U-Net-CH combined segmentation and classification task in one network, which can improve the performance of both tasks because of the commonalities across tasks.

Currently our approach does not involve axial and oblique radiographs; therefore, fractures that can only be identified from other angles are hard to be detected in the proposed system. In the future, the system can be extended to analyze not only lateral but also axial and oblique calcaneus radiographs. Giving more detailed labeling, the system can also be further developed to classify the fine-grained fracture type and fracture region type. In addition, efforts can be made to improve the segmentation and classification performance by implementing other backbones and segmentation architectures.

## 5. Conclusions

In this paper, we present an automatic analysis system of lateral calcaneus radiograph, which includes algorithms of anatomical angle measurement, fracture identification, and fracture region segmentation. The system handles the fickle rotation of calcaneus in radiographs by a Rotation-Invariant Regressive-Voting (RIRV) landmark detection method. For calcaneus fracture identification and segmentation, a CNN model named U-Net with auxiliary classification head (U-Net-CH) is designed. The input ROIs of the CNN are normalized by detected landmarks to uniform view, orientation, and scale. Our system is evaluated on a large calcaneus dataset and experimental results are promising, with 3.8 degree of BA mean angle error, 96.55% of classification

F1-score, and 0.586 of segmentation IoU-score. The whole calcaneus radiograph analysis system can serve as a fracture screening tool and a calcaneal angle measurement tool that aids the orthopedists' diagnosis. Currently, the algorithm has been packed and delivered to collaborating hospitals for clinical validation. In future work, we plan to include axial and oblique radiographs together with lateral calcaneus radiographs in our system to make the diagnosis a multi-mode task.

# Acknowledgments

We would like to thank our partner for this research project: Beijing Yiwave technology and First Affiliated Hospital of Jinzhou Medical University, for the radiograph data and labeling.

# Appendix

|  | $[\Delta\theta_{min}, \Delta\theta_{max}]$ | $[s_{min}, s_{max}]$ | $D_{SR}$ | $N$ |
|---|---|---|---|---|
| **Stage One(h=1)** | N/A | [30,50] | N/A | 100 |
| **Stage Two(h=2)** | N/A | [25,35] | 440 | 100 |
| **Stage Three(h=3)** | [-π/6, π/6] | [15,20] | 300 | 80 |
| **Stage Four(h=4)** | [-π/6, π/6] | [10,16] | 200 | 50 |

Table 5 Hyper parameters in RIRV training

|  | $[\Delta\theta_{min}, \Delta\theta_{max}]$ | $[s_{min}, s_{max}]$ | $D_{SR}$ | $N$ | $Th$ |
|---|---|---|---|---|---|
| **Stage One(h=1)** | N/A | [35,45] | N/A | 200 | 100 |
| **Stage Two(h=2)** | N/A | [25,35] | 300 | 100 | 60 |
| **Stage Three(h=3)** | [-π/12, π/12] | [16,19] | 160 | 80 | 30 |
| **Stage Four(h=4)** | [-π/12, π/12] | [13,14] | 80 | 50 | N/A |

Table 6 Hyper parameters in RIRV prediction

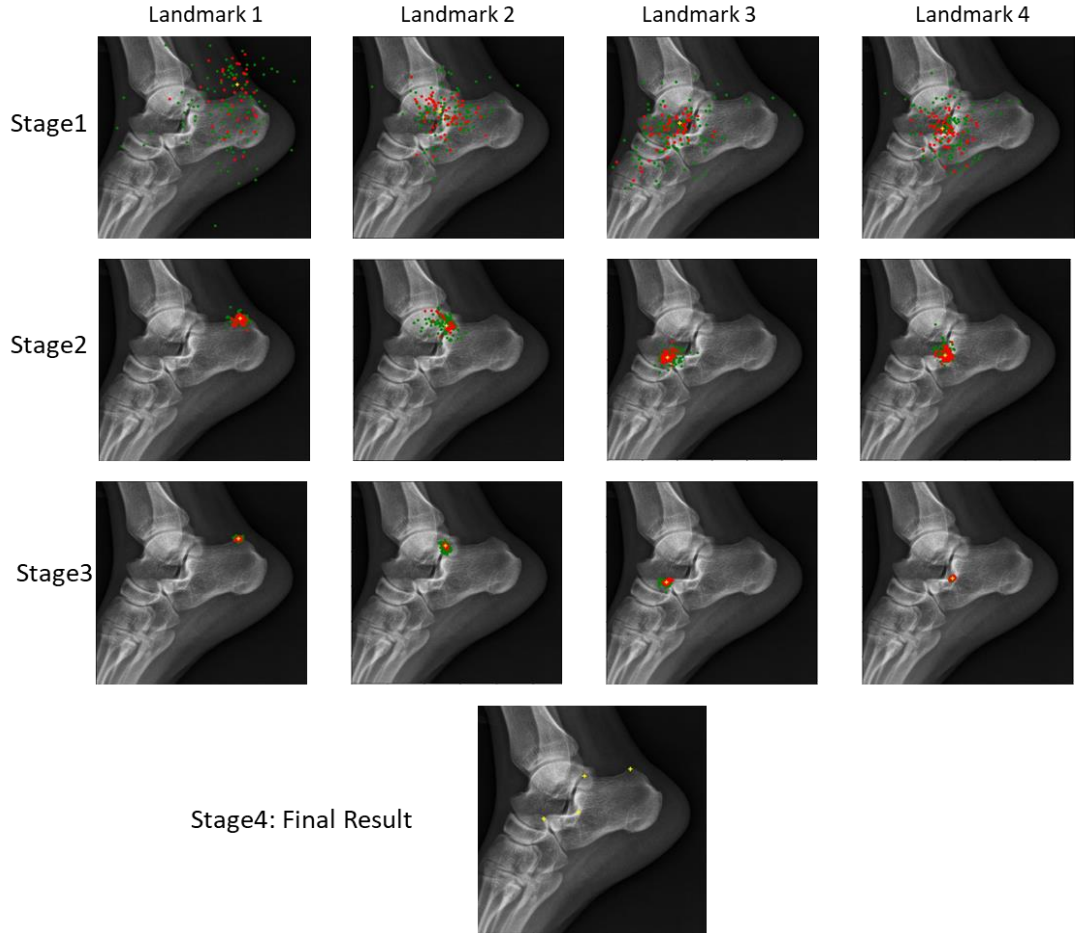

Figure 15 Results of each stage in RIRV. Original voting candidates **c** are denoted as green point and HPDV screened valid candidates $c^{valid}$ are denoted as red point.

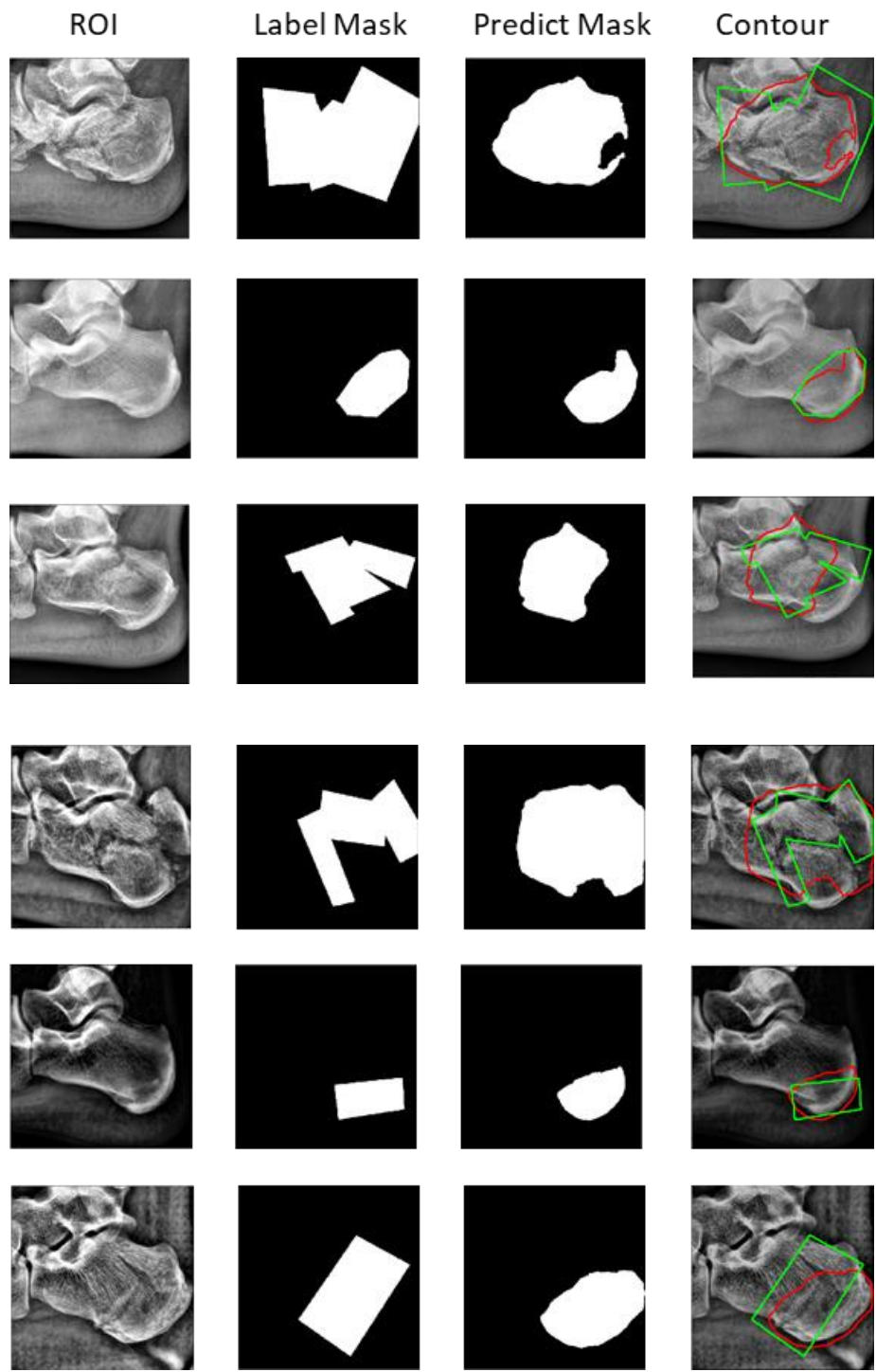

Figure 16 Results of fracture segmentation by U-Net-CH. The four columns are normalized calcaneus ROI, ground truth fracture region, fracture region predicted by U-Net-CH, the contour of fracture region, from left to right respectively. The green and red contour line denote labeled region and predicted region respectively.

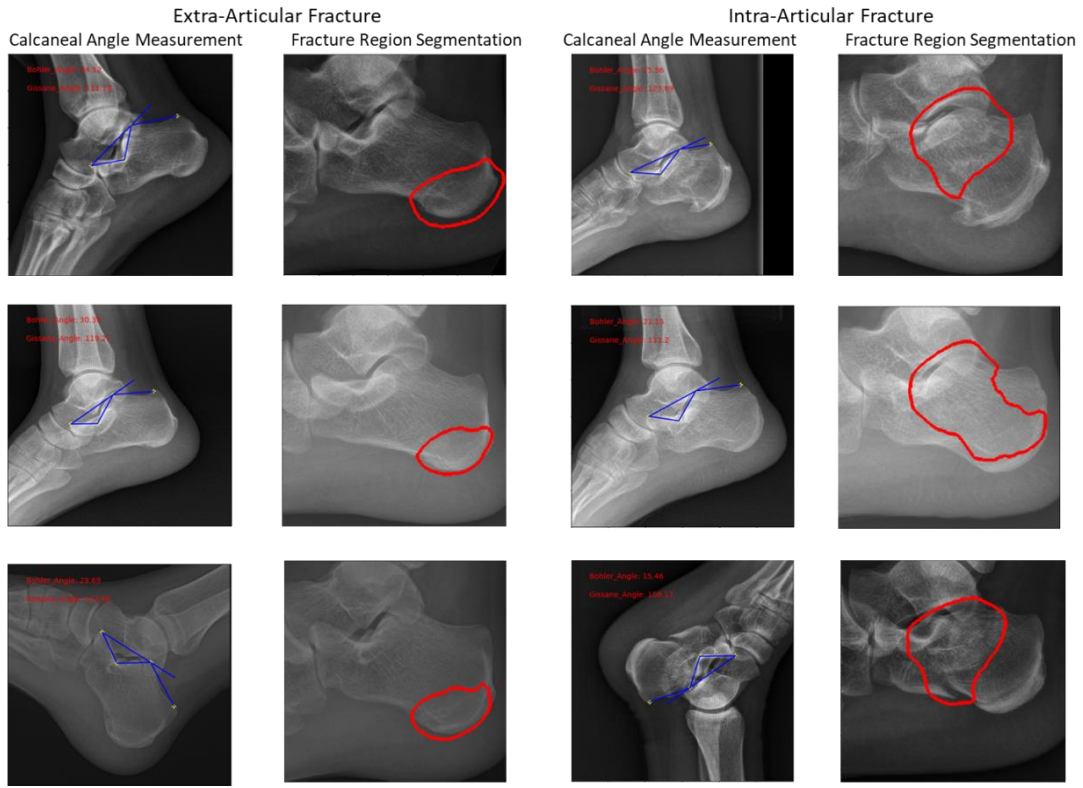

Figure 17 Results of calcaneus radiograph analysis system. The blue line denotes the side of BA and CAG. The red contour denoted the segmented fracture region.